# Faded-Experience Trust Region Policy Optimization for Model-Free Power Allocation in Interference Channel

Mohammad G. Khoshkholgh, *Member, IEEE,* and Halim Yanikomeroglu, *Fellow, IEEE*

*Abstract*—Policy gradient reinforcement learning techniques enable an agent to directly learn an optimal action policy through the interactions with the environment. Nevertheless, despite its advantages, it sometimes suffers from slow convergence speed. Inspired by human's decision making approach, we work toward enhancing its convergence speed by augmenting the agent to memorize and use the recently learned policies. We apply our method to the trust-region policy optimization (TRPO), primarily developed for locomotion tasks, and propose faded-experience (FE) TRPO. To substantiate its effectiveness, we adopt it to learn continuous power control in an interference channel when only noisy location information of devices is available. Results indicate that with FE-TRPO it is possible to almost double the learning speed compared to TRPO. Importantly, our method neither increases the learning complexity nor imposes performance loss.

## I. INTRODUCTION

RECENT YEARS have witnessed a surge in the use of machine learning techniques, see, e.g., [1], [2], for end-to-end model-free resource allocation in communication systems [3]–[5], which are useful for the cases that the model of the communication system is unknown or is not explicitly learned. This allows more robust, versatile, and scalable solutions for complex scenarios when the conventional approaches, based on well-crafted optimization problems, could fail or render substantial complexity.

In [4] authors trains a deep neural network (DNN) to learn power allocation in wireless interference channel. [5] introduces deep power control (DPC) via adopting the convolutional DNN [1] to train an interference channel for learning the optimal power control. Moreover, [6] develops power control net (PCNet) as an ensemble of convolutional DNN to deal with varying noise power in interference channel. [7] shows that by training the resource allocation problem in the dual domain it is possible to effectively account for the stochasitiy of the resource constraints.

On the other hand, one can use deep reinforcement learning (DRL) to deal with learning resource allocation via interactions with the communication medium and adjusting the actions. The use of deep Q-learning to derive the optimal power allocation in the cellular network has widely adopted [8]–[10]. The transmit powers however need to be firstly quantized, which is heuristic and could lead to performance degradation in large-dimensional action settings. In this paper, we exploit continuous DRL to tackle the power allocation [11]. In particular, we adopt policy gradient methods [12], [2], [13], which have achieved significant successes in challenging DRL problems such as robotics and gaming. Trust region policy optimization (TRPO) [13] is a modern policy gradient method that can achieve robust performance on a wide variety of challenging locomotion tasks. However, such typical policy gradient techniques may suffer from poor or unstable convergence, making them less sample efficient, which is disadvantages given that drawing samples in model-free RL settings is often expensive or time consuming [11]. To make TRPO more efficient for power allocation in wireless communications, in this work, we propose to further improve its sample efficiency by leveraging the humans' decision making tendency by using their past experiences along with the current gathered data. We call the developed algorithm faded experience TRPO (FE-TRPO) as it incorporates fading contribution of the past policies in updating new policy. We apply our method for complex problem of continuous power allocation in interference channel when the channel state information is not available at the transmitters (thus traditional techniques such as WMMSE are not applicable [14]). We show that compared to TRPO, FE-TRPO is able to nearly double the speed of learning without loosing performance and increasing the computational complexity, which makes policy gradient a more suitable solutions.

We should emphasize that our approach is different than imitation learning [15], learning from demonstration [16], and curriculum learning [17], whereby the main focus is to reuse/transfer the knowledge gathered from another setting often involving easier tasks. Circumventing the learning from the scratch, using such techniques an agent is feed with expert knowledge from other domains as an initial guide for enhancing the exploration performance. In contrast, our approach enables an agent to reuse its own learned policies for improving the policy faster. In effect, we enable the agent to both utilize the data and its own past experiences in the course of learning, which, to our best knowledge, has not discussed in the literature.

## II. A BRIEF BACKGROUND ON POLICY GRADIENT DRL

### A. Vanilla Policy Gradient

In continuous DRL an agent operating in an uncertain environment with the continuous state and action spaces interacts with the environment to learn an optimal policy [18]. In each interaction the agent takes an action $a_t \in \mathbb{R}^B$ ($B$ is the

The paper submitted August 04, 2020.
The authors are with the Department of Systems and Computer Engineering, Carleton University, Ottawa, Ontario, Canada. Email: m.g.khoshkholgh@gamil.com, halim @sce.carleton.ca.
This work was supported by Huawei Canada Co., Ltd.

action dimension) based on its observation of the environment state $s_t \in \mathbb{R}^S$ ($S$ is the dimension of the state space), which leads the agent to the new state $s_{t+1}$ along with receiving the bounded reward $r_t \in \mathbb{R}$. The policy guides the agent to what action should be taken in a certain state in order to maximize the reward via maximizing the aggregate (discounted) expected reward (or average return) [2]

$$J(\pi) = \mathbb{E}_\pi \sum_t \gamma^t r_t(s_t, a_t) \tag{1}$$

by finding an optimal policy $\pi_{\boldsymbol{\theta}}(a_t|s_t)$ (or for short $\pi_{\boldsymbol{\theta}}$) where $\boldsymbol{\theta}$ are the parameters of the associated DNN. Parameter $\gamma \in (0, 1]$ is the discount factor prioritizing short-term rewards and the expectation is on the policy $\pi$ as well as the stochastic environment dynamics. We focus on stochastic policies, i.e., $a_t \sim \pi_{\boldsymbol{\theta}}$, in which the DNN deterministically maps the state to a vector that specifies a distribution over the action space. To learn the policy we adopt policy gradient methods:

$$\nabla_{\boldsymbol{\theta}} J(\boldsymbol{\theta}) = \boldsymbol{g} = \mathbb{E}_{\pi_{\boldsymbol{\theta}}} \sum_t \nabla_{\boldsymbol{\theta}} \log \pi_{\boldsymbol{\theta}} A_{\boldsymbol{\theta}}(s_t, a_t). \tag{2}$$

Here we use the case that the policy gradient is formulated through the *advantage function* $A_{\boldsymbol{\theta}}(s_t, a_t)$, which is the subtraction of the Q-function[1] and state-value function: $A_{\boldsymbol{\theta}}(s_t, a_t) = Q_\pi(s_t, a_t) - V_\pi(s_t)$. Advantage function measures the relative advantage value of action $a_t$. Note that for many applications the advantage function is preferable to the Q-value function in order to estimate the gradient (2) as it yields a lower variance [19]. In practice, the above expectation should be estimated over a batch of data collected from the current policy via Monte Carlo (MC)[2] technique (sample based estimate of the policy gradient). The agent iteratively collects data $(s_t, a_t, r_t, s_{t+1})$, estimates the gradient of the policy, updates the policy, and then discards the data, which is the core of REINFORCE algorithm.

*B. Trust Region Policy Optimization (TRPO)*

To improve the stability, besides learning the policy, the value function needs to be learned [19]. This is the core concept of *actor-critic* technique in which an DNN—called the *actor* or the policy net—updates the policy while another DNN—called the *critic* or the value net—updates the value function's parameters denoted by $\phi$. The update can, for example, be via gradient ascent $\phi_{t+1} = \phi_t + \beta_\phi \delta_t \nabla_\phi V_\phi(s)$, where $\delta_t = r(s_t, a_t) + \gamma V_\phi(s_{t+1}) - V_\phi(s_t)$ is the correction, also known as temporal difference (TD) [2] and $\beta_\phi$ is the learning rate of the critic network.

In practice, even with using the actor-critic configuration the vanilla policy gradient techniques, such as REINFORCE algorithm, fall short to effectively tackle the high-dimensional state-action problems [20]. This is because the gradient ascent fails to take the steepest ascent direction in the metric of parameter space without *too much* divergence from the current policy. The TRPO algorithm [13], [20] addresses this issue via imposing Kullback-Leibler (KL) divergence[3] constraint on the size of policy update. Recalling that the policy is stochastic, KL divergence is a natural choice as it quantifies the closeness of two probability distributions. In TRPO, a *surrogate objective function* is considered, as an estimate of the average return $J(\pi_{\boldsymbol{\theta}})$ (1). In each iteration the following optimization problem needs to be solved for updating the policy net:

$$\mathcal{O}: \text{Maximize}_{\boldsymbol{\theta}} \quad \mathbb{E}_{\pi_{\boldsymbol{\theta}_k}} \left[ \frac{\pi_{\boldsymbol{\theta}}(a|s)}{\pi_{\boldsymbol{\theta}_k}(a|s)} A_{\boldsymbol{\theta}_k}(s, a) \right] \tag{3}$$

$$\text{s.t.} \quad \mathbb{E}_{s \sim \pi_{\boldsymbol{\theta}_k}} [D_{KL}(\pi_{\boldsymbol{\theta}_k}(.|s) || \pi_{\boldsymbol{\theta}}(.|s))] \leq \delta_{KL}. \tag{4}$$

For a detailed discussion on the relationship between the surrogate objective function and actual expected return (1) refer to [13]. In short, this optimization problem attempts to update the current policy $\pi_{\boldsymbol{\theta}_k}$ via finding a new (relatively close) policy $\pi_{\boldsymbol{\theta}}$ by maximizing an scaled advantage function. The constraint, which is called *trust region constraint*, is KL divergence constraint between the current policy and the new policy. In this form the optimization problem $\mathcal{O}$ is not computationally affordable, hence an approximate optimization problem is then considered instead:

$$\tilde{\mathcal{O}}: \text{Maximize}_{\boldsymbol{\theta}} \quad \hat{\boldsymbol{g}}^T (\boldsymbol{\theta} - \boldsymbol{\theta}_k) \tag{5}$$

$$\text{s.t.} \quad (\boldsymbol{\theta} - \boldsymbol{\theta}_k)^T \hat{F}_{\boldsymbol{\theta}_k} (\boldsymbol{\theta} - \boldsymbol{\theta}_k) \leq \delta_{KL}. \tag{6}$$

where the objective function is the first-order approximation of the surrogate objective function and the constraint is the second-order approximation of the KL divergence constraint (4). Here $\hat{g}$ is the estimate policy gradient and $\hat{F}_{\boldsymbol{\theta}_k}$ is the estimate Fisher information matrix (FIM) associated to the average KL divergence at the current policy $\boldsymbol{\theta}_k$ [13].

III. FADED-EXPERIENCE TRPO (FE-TRPO)

In TRPO, as all the policy gradient methods, the agent attempts to update the policy in each iterations merely by exploiting the collected data. Although TRPO is able to improve the sample efficiency substantially, there could be room to increase its convergence speed. This is the focus and the main contribution of the paper. We incorporate the steps that humans (naturally) follow for making decisions: 1) while we gather new data/experience/knowledge to cope with emerging situations, we simultively tend to exploit the past (relevant) experiences; 2) To make the positive (cors. negative) outcome(s) more (cors. less) probable, we intuitively tend to weigh more on most recent experiences over the distant ones; 3) Regardless of the relevancy of the past experiences, we tend to relay more on the current situation/data. Refer to algorithm (1) for the pseudo-code of FE-TRPO.

In what follows, we go through main steps of FE-TRPO algorithm and highlight its difference with TRPO.

---

[1] For given policy $\pi$, the *state-value function* $V^\pi(s_t)$ measures the expected discounted reward from state $s_t$ via $V^\pi(s_t) = \mathbb{E}_{a_t, s_{t+1}, \ldots} \sum_{t' \geq t} \gamma^{t'-t} r_{t'}(s_{t'}, a_{t'})$. The *Q-function* is similarly defined as $Q^\pi(s_t, a_t) = \mathbb{E}_{s_{t+1}, a_{t+1}, \ldots} \sum_{t' \geq t} \gamma^{t'-t} r_{t'}(s_{t'}, a_{t'})$, which is the state-value function for a given action.

[2] We use the hat symbol to highlight that the quantity is the sample estimate of the mathematical expectation. Thus the estimated gradient of (2) is denoted by $\hat{g}$.

[3] For probability distributions $P$ and $Q$ over a given random variable the KL divergence is defined as $D_{KL}(P||Q) = \mathbb{E}_P[\log \frac{P}{Q}]$.

As TRPO, FE-TRPO has an outer loop indexed by $l = 1, 2, \ldots, L$. For each iteration $l$, the policy is fixed. The iteration comprises of an inner loop allowing the agent collect $N$ transitions, which also known as batch size. Note that each $T$ consecutive transition is an episode. Using the collected transitions the advantage function, gradient, and FIM are estimated via MC technique, which are used to update the policy network and value network.

*Policy:* From Step (5), the policy in each iteration $l$ is chosen to be a combination of the current policy $\pi_l(\boldsymbol{\theta}_l)$ and $M$ past policies $\pi_{\min(l-m,0)}(\boldsymbol{\theta}^{(m)}_{\min(l-m,0)})$ as

$$\tilde{\pi}_l = w_0 \pi_l(\boldsymbol{\theta}_l) + \sum_{m=1}^{M} w_m \pi_{\min(l-m,0)}(\boldsymbol{\theta}^{(m)}_{\min(l-m,0)}), \quad (7)$$

where $\boldsymbol{\theta}^{(m)}_{\min(l-m,0)}$ are the parameters of past experience $m$. Weights are given and kept fixed during the course of learning. One is free to choose any possible combinations as far as they stay in the set $\mathcal{W} = \{w_m : w_0 \in (0,1], w_0 \geq w_m, w_m \in [0,1], m \in [1,M], \sum_{m=0}^{M} w_m = 1\}$. In short, the highest weight is assigned to the current policy. Note that all policies can be weighted equally, however, it is generally intuitive to emphasize more on the more recent policies. It is important to note that as the past policies $\pi_{\min(l-m,0)}(\boldsymbol{\theta}^{(m)}_{\min(l-m,0)})$ are already learned and memorized, the agent does not overwhelmed by any extra computational cost. Moreover, by choosing $M = 1$, FE-TRPO reduces to TRPO.

*Updating Policy:* Updating policy is similar to the TRPO and carried out based on solving optimization problem $\tilde{\mathcal{O}}$ in several steps (Step (6) to Step (11)). First, we need to estimate the rewards-to-go $\widehat{\boldsymbol{R}}$ and advantages $\widehat{\boldsymbol{A}}$. In (8), $d_t \in \{0,1\}$, where $d_t = 1$ implies that the episode is terminal. As a result, the reward of the terminated time step of the episode is not included in calculation of the advantages and rewards-to-go. On the other hand, in the calculation of the advantages $\widehat{\boldsymbol{A}}$ we adopt the generalized advantage estimation (GAE) [19] where $\lambda \in (0,1]$ is a given parameter to improve the stability.

The estimated advantages are then used to estimate the gradient over the batch in Step (9). Steps (10) and (11) are to take the maximum step for updating the current policy. First, in Step (10) we derive a new direction via the conjugate gradient algorithm in order to efficiently solve $\widehat{F}_{\boldsymbol{\theta}_k} \hat{\boldsymbol{x}}_k = \hat{\boldsymbol{g}}$ through several iterations instead of resorting to the computation of the inverse of FIM. This substantially increases the computation efficiency and memory usage as the underlying DNN could have millions of parameters. Step (11), known as line search in TRPO algorithm, is a crucial step as it ensures that the new policy—which is derived based on the approximations of the objective and the constraint—guarantees that the actual surrogate objective (not its linear approximation) is improved while the Kl divergence constraint (not its quadratic approximation) stays fulfilled. In effect, by the line search the largest legitimate step toward the next policy is taken. For a given backtracking coefficient $\alpha < 1$ the parameters $\boldsymbol{\theta}_l$ are updated up to the maximum backtracking steps $J$. We terminate the line search when the smallest value $\alpha^j$ (the bigger is $j$, the

**Algorithm 1** Faded-Experience TRPO (FE-TRPO)

1: Hyper-parameters: KL divergence limit $\delta_{KL}$, backtracking coefficient $\alpha$, maximum number of backtracking steps $n_B$, behavioral memory size $M$, GAE lambda $\lambda \in (0,1]$, number of transitions $N$
2: Input: initialize policy parameters $\boldsymbol{\theta}_0$, initial value function parameters $\boldsymbol{\phi}_0$
3: Initialize the memorized polices parameters $\boldsymbol{\theta}^{(m)}_0$ with $\boldsymbol{\theta}_0$ and behavioral weights from $\mathcal{W} = \{w_m : w_0 \in (0,1], w_0 \geq w_m, w_m \in [0,1], m \in [1,M], \sum_{m=0}^{M} w_m = 1\}$
4: **for** $k = 0, 1, 2, \ldots L$ **do**
5:     Collect $N$ transitions $(\boldsymbol{s}_t, \boldsymbol{a}_t, r_t, \boldsymbol{s}_{t+1})$ by running policy

$$\tilde{\pi}_k = w_0 \pi_k(\boldsymbol{\theta}_k) + \sum_{m=1}^{M} w_m \pi_{\min(k-m,0)}(\boldsymbol{\theta}^{(m)}_{\min(k-m,0)})$$

6:     Set $\widehat{\boldsymbol{R}} = \boldsymbol{0}$ and $\widehat{\boldsymbol{A}} = \boldsymbol{0}$
7:     **for** $t = N-1, \ldots, 1, 0$ **do**

$$\begin{cases} \widehat{\boldsymbol{R}}[t] = r_t + \gamma(1-d_t)\widehat{\boldsymbol{R}}[t+1] \\ \hat{\delta} = r_t + \gamma(1-d_t)V_\phi(\boldsymbol{s}_{t+1}) - V_\phi(\boldsymbol{s}_t) \\ \widehat{\boldsymbol{A}}[t] = \hat{\delta} + \gamma\lambda(1-d_t)\widehat{\boldsymbol{A}}[t+1] \end{cases} \quad (8)$$

8:     **end for**
9:     Estimate the policy gradient

$$\hat{g} = \frac{1}{N}\sum_{t=0}^{N-1} \nabla_{\boldsymbol{\theta}_k} \log \tilde{\pi}_k \widehat{\boldsymbol{A}}[t], \quad (9)$$

10:     Use the conjugate gradient algorithm to compute $\hat{F}_k \hat{\boldsymbol{x}}_k = \hat{\boldsymbol{g}}$
11:     Update the policy parameters:

$$\boldsymbol{\theta}_{k+1} = \boldsymbol{\theta}_k + \alpha^j \sqrt{\frac{2\delta_{KL}}{\hat{\boldsymbol{x}}_k^T \hat{F}_{\boldsymbol{\theta}_k}^{-1}\hat{\boldsymbol{x}}_k}} \hat{\boldsymbol{x}}_k, \quad j = \{0, 1, 2, \ldots, K\} \quad (10)$$

12:     **for** $m = M, \ldots, 2, 1$ **do**
13:         $\boldsymbol{\theta}^{(m)}_k = \boldsymbol{\theta}_k \mathbf{1}_{\{m=1\}} + \boldsymbol{\theta}^{(m-1)}_k \mathbf{1}_{\{m>1\}}$
14:     **end for**
15:     Update the value function

$$\boldsymbol{\phi}_{k+1} = \operatorname{argmin}_\phi \frac{1}{N}\sum_{t=0}^{N-1}\left(V_\phi(\boldsymbol{s}_t) - \widehat{\boldsymbol{R}}[t]\right)^2. \quad (11)$$

16: **end for**

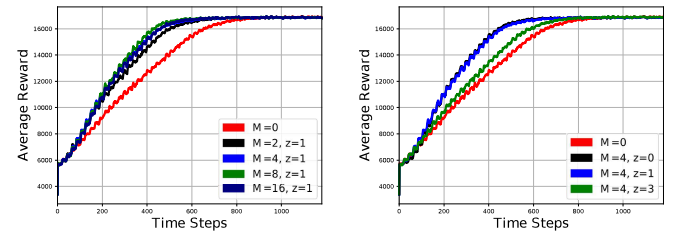

Fig. 1. (a) Average reward versus time steps for different values of $M$. (b) Average reward versus time steps for different values of $z$.

smaller will be the update step) satisfies the KL divergence constraint.

*Updating Experiences:* After learning new policy we need to update the past experiences. In essence, we need to register the current policy in iteration $l$ as the most recent memorized policy, and shift the rest of the experiences one step backward, meaning that the agent dumps the last experience. This is done in Steps (12)-(14). This way, the agent is able to always keep the memory up-to-date.

*Value Network:* The update of the value network $V_{\boldsymbol{\phi}_k}$ is done in Step (15) via the supervised learning method using the rewards-to-go $\widehat{\boldsymbol{R}}$ and mean-squared-error regression.

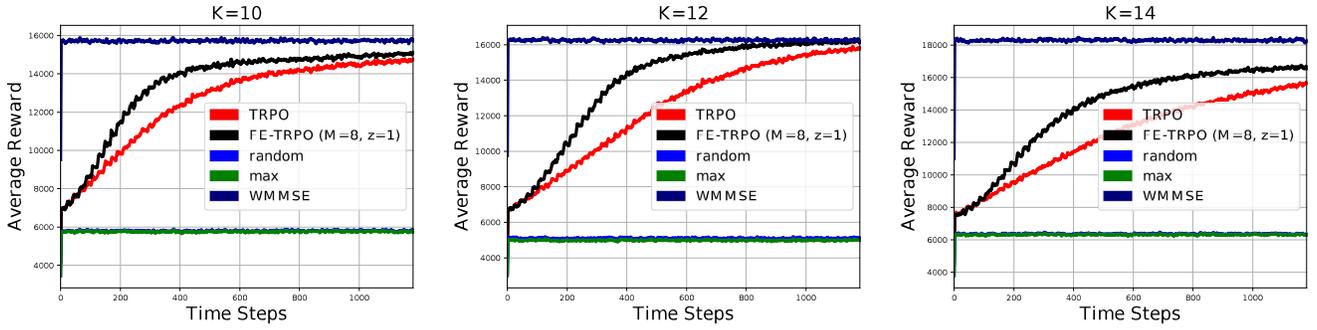

Fig. 2. Average reward versus time steps for different values of $K$.

## IV. POWER CONTROL IN INTERFERENCE CHANNEL

To measure the performance gain of FE-TRPO we here use it to learn the continuous power control in an interference channel.

*1) Problem Formulation:* The interference channel consists of $K$ single-antenna transceivers, where each transmitter $k$ has its own intended receiver with $h_{kk} \in \mathbb{R}^+$ standing as the corresponding channel power gain. Transmitter $k$ poses interference on the other receivers through channel power gains $h_{kj} \in \mathbb{R}^+$. We assume the interference is considered as noise. Signal-to-interference-plus-noise (SINR) at the receiver $k$ is $\text{SINR}_k = \frac{h_{kk} P_k}{\sigma_k^2 + \sum_{j \neq k} h_{jk} P_j}$ where $\sigma_k^2$ is the noise power at the receiver $k$ and $P_k \in [0, \hat{P}_k]$ is the (continuous) transmission power, which should be smaller than the maximum permissible transmission power $\hat{P}_k$. For training we set $\hat{P}_k = 1$ W, $\forall k$. The data rate of user $k$ is calculated by the Shannon's formula $r_k = \log(1 + \text{SINR}_k)$.

We consider a circular area with radius 60 m, and randomly locate users in it. The wireless channel is based on 3GPP Line-of-Sight (LOS)/none-LOS (NLOS) path-loss attenuation model $L_{jk} = \|X_{jk}\|^{-\alpha_l} \sim p_l(\|X_{jk}\|)$ for $l \in \{L, N\}$ [22], where $p_N(\|X_{jk}\|) = 1 - p_L(\|X_{jk}\|)$ is the probability of LOS that is a function of distance $\|X_{jk}\|$: $p_L(\|X_{jk}\|) = \min\left\{\frac{D_0}{\|X_{jk}\|}, 1\right\}\left(1 - e^{-\frac{\|X_{jk}\|}{D_1}}\right) + e^{-\frac{\|X_{jk}\|}{D_1}}$, which is also known as ITU-R UMi model. Also, $\alpha_L$ (resp. $\alpha_N$) is the path-loss exponent associated with LOS (resp. NLOS) component where $\alpha_N > \alpha_L$. $D_1$ and $D_0$ are hyper-parameters which can take different values for different environments. We set the channel parameters as $\alpha_L = 2.4$, $\alpha_N = 3.78$, $D_0 = 18$ m, $D_1 = 36$ m, and the background noise power $-173$ dBm/Hz. The fading power gain under the LOS mode is modelled by Nakagami-m distribution with parameter $m = 10$. Under the NLOS mode the fading is modelled via unit-mean exponential random variable. We also consider large-scale shadowing with mean zero dB and standard deviation 5 dB under LOS mode and 8.6 dB under NLOS mode. Receivers and transmitters are allowed to dislocate by up to 5 meters in a random direction at the start of each iteration. However, we are making sure that the receivers stay in the simulation area. The optimal power allocation can be obtained by solving the following optimization problem:

$$\max_{P_1, \ldots, P_K} \sum_k \log(1 + \text{SINR}_k) \quad \text{s.t.} \quad 0 \leq P_k \leq \hat{P}_k, \quad \forall k.$$

A nearly optimal solution can be obtained via WMMSE algorithm [14] provided that CSI is perfectly known at the transmitters as well as the receivers. However, when the CSI is not known perfectly, the model-free solutions based on DRL is recommendable. We assume that a perturbed distance information $\|X_{jk}\|(1 + \Delta_{jk})$ is known at the transmitters, where $\Delta_{jk} \in [0.9, 1.1]$. We like to allocate power merely based on the perturbed distance information in order to maximize the sum rate. We compare the performance with that of WMMSE (which relies upon the accurate CSI) as well as random power allocation and the case of $P_k = \hat{P}_k$.

*2) Configuration:* As the state and action spaces are continuous it is a proper choice to adopt the multivariate normal distribution with diagonal covariance matrix, i.e., $\pi_{\boldsymbol{\theta}} \sim \mathcal{N}(\text{mean}_{\boldsymbol{\theta}}, \text{diag}(\boldsymbol{v}))$. By the aid of a DNN with dense layers the features are extracted and the state space mapped into the mean of the Gaussian distribution [20], [13]. The logarithm of the associated standard deviation has not have its own set of parameters, thus is computed from the DNN's head—no need to consider a separate DNN to map the features into the standard deviation. Mathematically, given a DNN with $Q$ dense layers with weights and biases $\{\boldsymbol{W}_i, \boldsymbol{b}_i\}_{i=1}^Q$ and a vector $\boldsymbol{u} \in \mathbb{R}^B$, the policy is formulated via $\mathcal{N}(\text{DNN}(\boldsymbol{s}, \{\boldsymbol{W}_i, \boldsymbol{b}_i\}_{i=1}^Q), e^{\boldsymbol{u}})$. In this model $\boldsymbol{u}$ is the logarithm of the standard deviation of the Gaussian policy. Therefore, the closed form expression for the KL divergence is readily derivable:

$$D_{KL}(\mathcal{N}(\text{mean}_{\boldsymbol{\theta}}, \text{diag}(\boldsymbol{v})) \| \mathcal{N}(\text{mean}_{\bar{\boldsymbol{\theta}}}, \text{diag}(\boldsymbol{w}))) =$$

$$\sum_i \left(w_i^2 - v_i^2 + \frac{v_i^2 + ([\text{mean}_{\boldsymbol{\theta}}]_i - [\text{mean}_{\bar{\boldsymbol{\theta}}}]_i)^2}{2w_i^2} - 0.5\right),$$

which simplifies the evaluation of the FIM.

The mean of this distribution is a DNN with 3 dens layers. The first and second layers are with input/output dimensions $S/400$ and $400/300$ respectively, where $S$ is the space dimension. This DNN has two heads, one for the mean value and the other for the logarithm of the standard deviation. Each of these are modelled by its associated dense layer with size $300/K$. Similarly, the value net is also a DNN with three layers with the difference that the last layer has dimensions $300/1$. The activation functions are Tanh [1]. In this experiment we configure the state space as the perturbed distance information from all users along with the transmitted data rates. At each iteration we let the state keep three previous channel realizations too. Thus $S = 4(K^2 + K)$.

Regarding FE-TRPO, we set the weights based on $w_i = \frac{1}{(i+1)^z}$ for $i = 0, 1, \ldots, M$, where $i = 0$ is associated with the current policy. As seen, by increasing $z$ the emphasis on the past experiences is reduced.

For the experiments we use the pytorch library [21]. For each experiment we consider 6 different random seeds and calculate the average values accordingly. In our experiment we set $L = 1300$, $N = 10000$ with episode length $T = 500$. Furthermore, we set $\delta_{KL} = 0.05$, and GAE $\lambda = 0.94$.

*3) Performance Evaluation:* In Fig. 1 we compare the performances of FE-TRPO and TRPO. In Fig. 1-(a) we set $z = 1$ and study the impact of $M$. As seen, by growing $M$ the convergence speed of FE-TRPO increases. For $M = 8$ the convergence speed of FE-TRPO (almost) doubles compared to TRPO, which is an impressive gain noticing the simplicity of FE-TRPO. In Fig. 1-(b) we keep $M = 4$ and study the impact of $z$. We observe that when $z = 3$ the performance gain of FE-TRPO is marginal, however for $z = 1$ it is possible to double the convergence speed.

In Fig. 2, we study the convergence speed and performance of TRPO and FE-TRPO for several choices of $K$ (the number of users). We also compare the reward with that of WMMSE, random power allocation, and the maximum power allocation. As seen, although TRPO and FE-TRPO only have access to the perturbed distance information they can (almost) achieve as does WMMSE, which requires perfect CSI. (Note that it the figures the performance of random power allocation and maximum power allocation coincides closely.) This highlights the impressive importance of model-free DRL for resource allocation. On the other hand, we observe that for all cases FE-TRPO is able to substantially expedite learning in comparison with TRPO. This makes the FE-TRPO a suitable choice for resource allocation in communication systems.

## V. Conclusions

We proposed FE-TRPO for expediting the learning speed of TRPO by allowing the agent memorizes and uses the past learned policies. We adopted it to learn continuous power control in an interference channel when the transmitters have access merely to the noisy distance information in order to demonstrate its effectiveness. Results indicate that using FE-TRPO it is possible to almost double the learning speed of TRPO and nearly achieve the optimal sum rate performance. We also discussed the impact of the number of past policies and their contributions on improving learning speed. Our method does not increase the learning complexity.